\documentclass[11pt]{article}          % LaTeX 2e 
\usepackage{times}                               % LaTeX 2e 
\usepackage{graphicx,epsfig,moreverb,multirow,indentfirst,latex8}
\usepackage{subfigure}
\usepackage{epstopdf}

\begin{document}
%{\small\em ISCA-2003 Submission.  Do not distribute.}
\thispagestyle{empty}
\setlength{\baselineskip}{1.5\baselineskip}

% Hand-made title page
\begin{centering}

\begin{minipage}{0.0in}
\vspace{1.5in}
\end{minipage}

\begin{Large}
\begin{bf}
Dynamic Merge Point Prediction
\end{bf}

\vspace{0.5in}
\begin{em}
Stephen Pruett and Yale Patt
\end{em}

\end{Large}

\vspace{0.75in}
\begin{figure}[h]
\centerline{\includegraphics[width=2.0in]{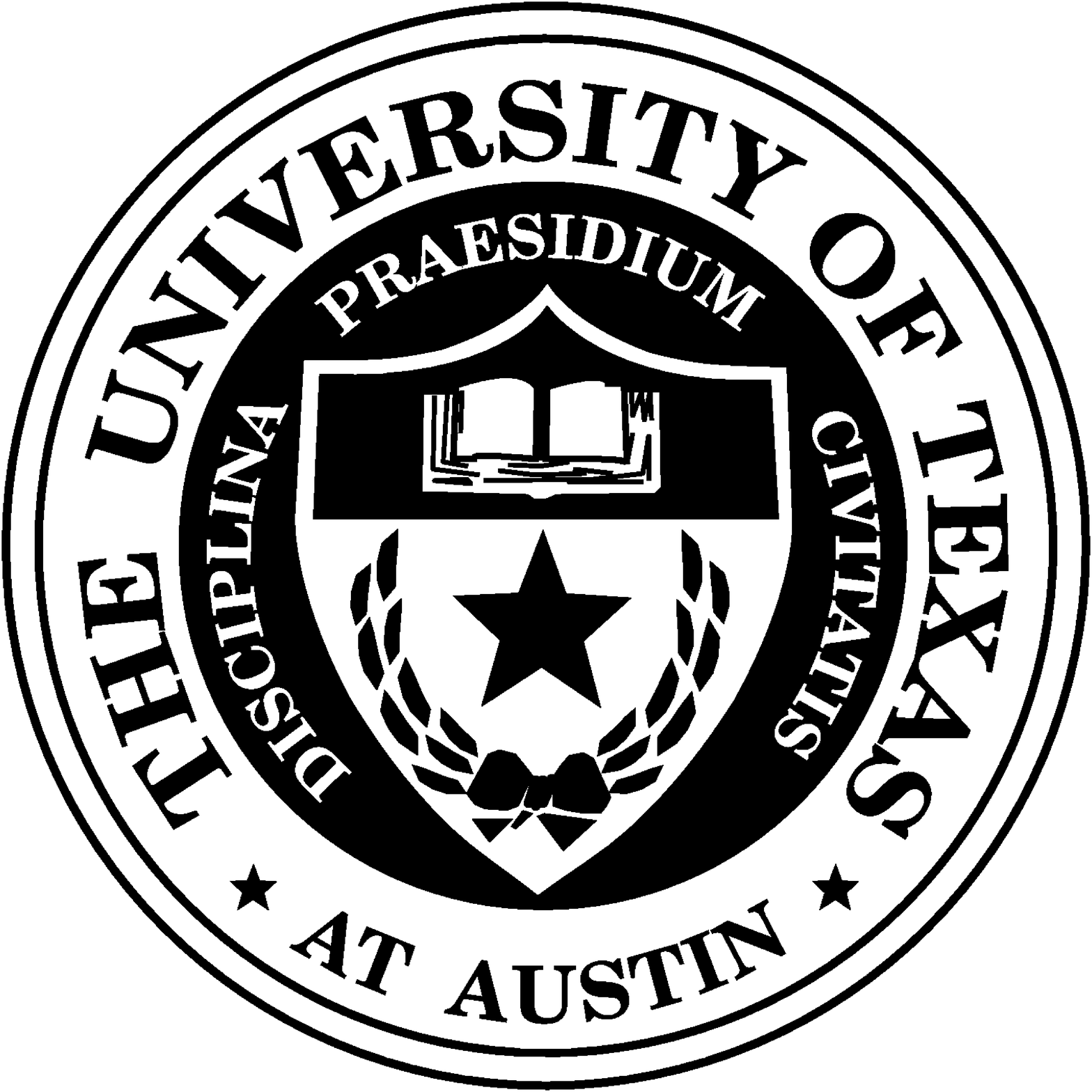}}
\vspace{0.75in}
\end{figure}

\begin{bf}
\begin{Large}
High Performance Systems Group \\
\end{Large}
\begin{large}
Department of Electrical and Computer Engineering \\
The University of Texas at Austin \\
Austin, Texas 78712-0240 \\
\end{large}
\end{bf}

\vspace{1.22in}
\begin{large}
\begin{bf}
TR-HPS-2020-001 \\
April, 2020 \\
\end{bf}
\end{large}

\end{centering}
\pagebreak

\pagenumbering{arabic}

\title{Dynamic Merge Point Prediction}

\author{
{\begin{tabular}[t]{ccc} Stephen Pruett  & Yale Patt  \end{tabular}}
\\
\\
\begin{normalsize}
\begin{minipage}{3.0in}
\begin{centering}
Electrical and Computer Engineering \\
The University of Texas at Austin \\
stephen.pruett@utexas.edu, patt@ece.utexas.edu \\
\end{centering}
\end{minipage}
\end{normalsize}
}
\maketitle

\begin{abstract}
  Despite decades of research, conditional branch mispredictions still pose a significant
  problem for performance. Moreover, limit studies on infinite size predictors show that
  many of the remaining branches are impossible to predict by current strategies.
  Our work focuses on mitigating performance loss in the face of impossible to predict
  branches. This paper presents a dynamic merge point predictor, which uses instructions
  fetched on the wrong path of the branch to dynamically detect the merge point. Our
  predictor locates the merge point with an accuracy of 95\%, even when faced with
  branches whose direction is impossible to predict. Furthermore, we introduce a
  novel confidence-cost system, which identifies costly hard-to-predict branches. Our
  complete system replaces 58\% of all branch mispredictions with a correct merge point
  prediction, effectively reducing MPKI by 43\%. This result demonstrates the potential
  for dynamic merge point prediction to significantly improve performance.
\end{abstract}

\newcommand{\slewofcite}{\cite{Cher, Rotenberg, Hilton, roth2002squash, Sodani, Meng, Chou,
SYRANT, SPREPI, Al-Zawawi, TraceProcessors}}

\section{Introduction} \label{sec:intro}

Branch prediction is a fundamental part of all high-performance microarchitectures. High accuracy is required
to maintain the high fetch bandwidth demanded by the running program. Unfortunately, there are many branches,
such as data-dependent branches, that are considered impossible-to-predict. In these cases, branch prediction
will always fall short. Despite this, branch prediction remains the only runtime solution for conditional
branches.

This paper discusses dynamic merge point prediction as a runtime alternative to branch prediction. By predicting the
merge point of a branch, the processor can avoid an expensive branch misprediction, instead utilizing a
control independence strategy \slewofcite{}. A control independence strategy is a technique that does not require knowledge of the branch direction, but can be used to mitigate or avoid a branch misprediction.
This paper proposes and evaluates a fundamentally new algorithm for detecting merge points. Prior approaches
use compiler heuristics and assumptions about code layout to predict the location of the merge point.
Our work takes advantage of branch mispredictions by comparing instructions fetched from the wrong path and correct path to detect the merge point. We argue this new approach to merge point prediction is fundamentally more
accurate and reliable than prior work. 

Merge point prediction is not intended to replace branch prediction, but rather supplement it when the
branch predictor has low confidence. Ideally, we would use the branch predictor whenever the possibility of
a misprediction is low, and use the merge point predictor when a branch misprediction is likely. To this end, 
we introduce a novel confidence-cost metric that identifies situations where branch prediction is too risky.
Our confidence cost predictor not only estimates the confidence of the branch, but also the expected latency 
for the branch to resolve.

Our merge point predictor is able to achieve an
average accuracy of 95\% across the SPEC CPU2006 benchmark suite
\cite{Spec}. The improved accuracy results in successfully detecting and
replacing 58\% of all branch mispredictions with a correct merge point
prediction, reducing the MPKI by an average of 43\%. 
\begin{figure}[t!]
  \centering
  \includegraphics[width=2.5in] {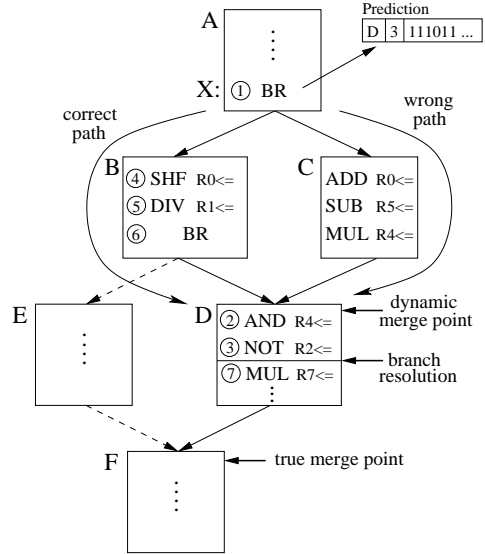}
  \caption{Example Control Flow Graph (CFG)}
  \label{fig:cfg}
  \medskip
  \raggedright
  \small
  \noindent
  \textbf{Dashed edges indicates paths that are rarely traversed at runtime.
  The compiler would report that the merge point of A is F. However, because
  block E is rarely seen, D is predicted as the merge point. }
\end{figure}

\section{Motivation} \label{sec:motivation}

The merge point predictor is designed with common control independence techniques in mind \slewofcite{}. 
Our work emphasizes three key principles that we believe to be essential for
utilizing control independence effectively. First, only hard-to-predict or
long latency branches are candidates for merge point prediction. We only consider merge point prediction an option when a branch misprediction is too risky. 
To achieve this we introduce a novel confidence-cost predictor, that considers both
the frequency of mispredictions and branch latency to estimate the total penalty of a mispredicted branch. These metrics are used together to identify branches
where the risk of misprediction is too great.
Second, predicted merge points should be as close to the branch as possible.
Often, the merge point predictor identifies more than one potential merge point for
a given branch. This is because the merge point predictor is identifying \textit{dynamic} merge points, which will be explained later in this section. Selecting merge points that are closer to the branch increases the number of post merge point instructions that are data-independent of the branch. Furthermore, it decreases the number of resources that are required by the merge point, which reduces the size of reservations required by some control independence strategies. 
Third, merge point predictions must be accurate. If a merge point prediction is wrong, then the machine must be flushed, similar to a branch misprediction. We design a highly accurate
dynamic merge point predictor that generates predictions at runtime without
relying on compiler input or code layout\footnote{Arrangement of basic
blocks in memory}. Prior predictors \cite{Cher, Collins} make assumptions
about compilers and code layout, making their work inaccurate and resistant
to change. Our predictor takes advantage of branch mispredictions, finding
the point where correct-path and wrong-path converge, making our predictor
oblivious to compiler changes.

\subsection{Weaknesses of Detecting Merge Points at Compile Time}
The compiler itself could be used to easily identify merge points with 100\% accuracy, however highly biased branches can weaken the compiler’s ability to find the nearest merge point~\cite{Diverge-Merge,Collins}, which can negatively affect performance. Furthermore, identifying hard-to-predict branches at compile time is difficult, reducing the compilers ability to provide help where it is most needed. Finally, compilers require costly instruction-set support to communicate with the microarchitecture that would likely result in additional fetch bandwidth being wasted.

Our dynamic merge point predictor uses run-time information to find the
nearest merge point. For example, consider the control-flow graph (CFG)
shown in Figure \ref{fig:cfg}. A compiler would identify block F as the
merge point, because it is the only block guaranteed to execute after A.
However, highly biased branches can effectively remove edges from the CFG.
To illustrate this, Figure \ref{fig:cfg} uses dashed edges to identify
branch directions that are rarely taken at run-time. If these edges are
omitted, then block D becomes the merge point. Predicting block D as the
merge point yields a merge point that is closer to the branch, but is
sometimes inaccurate. Predicting block F is always correct, but is farther
away than block D, making it less useful for performance.

Highly skewed branches prune edges of the CFG, producing merge points that
are closer to branches. 
Our experiments show that 61\% of
conditional branches never change direction while an additional 9\% of
branches change direction $<$1\% of the time \footnote{Measured across the
SPEC CPU2006 \cite{Spec} benchmark suite}. The large number of highly
biased branches suggests that identifying merge points at runtime will have
a major advantage over compile time. 

\subsection{Weaknesses of Prior Work in Merge Point Prediction}
The previous state-of-the-art merge predictor proposed by Collins et al.~\cite{Collins} has several major weaknesses. First, their predictor is not a general solution. It is a collection of three heuristics that all rely on the compiler to generate code that fits into their model. As compilers change over time, their predictor may become less accurate. In contrast, our algorithm leverages branch mispredictions to find the place where the wrong path and the correct path overlap. We do not rely on compiler heuristics, which enables us to cover a lot more cases and achieve higher accuracy. In our experiments, we compare to the infinitely sized, unrealistic predictor proposed by Collins et al. Despite their model having an unrealistic storage budget, their model achieves an average accuracy of only 78\% across the branch intensive workloads in SPEC 2006. Our realistic 4KB predictor achieves an accuracy of 95\% on those same workloads.

These numbers do not match the numbers reported by Collins et al. In their paper~\cite{Collins}, the authors report an accuracy of 95\% for their infinitely sized predictor, however, our evaluation shows an accuracy of at most 78\%.  We have accounted for the discrepancy and attribute it to two factors.  First, we do not account for branches with trivial merge points that are are unlikely to be mispredicted. Examples of this are  loop branches and function calls. In both cases, the merge point is trivial to predict, boosting the accuracy of the merge predictor. However, in both cases the branch direction is also trivial to predict, meaning that there will likely not be a branch direction misprediction. If the branch predictor is correct, we will not make use of the predicted merge point, making the correct merge point prediction meaningless. We therefore do not count loop branches and function calls as part of accuracy. In our system, only branches with low branch prediction confidence make use of the merge point predictor. Due to the high frequency of loop branches, removing them from consideration significantly lowers accuracy.  Second, we enforce that all merge points identified by both predictors be points where control converges. Due to the methodology used in~\cite{Collins}, some of the predicted values are not true merge points, but rather random intermediate places in the control flow graph. In our methodology, these points are counted as incorrect for both predictors.
 % 1 page
\section{Dynamic Merge Point Prediction} \label{sec:mp}

A merge point prediction consists of three parts: the PC of the merge point,
the merge distance, and the independent register set. The merge distance is
the predicted number of dynamic instructions in which the merge point is
expected to be found.  The predicted distance can be used to identify merge
point mispredictions, and also serves to
place an upper bound on the number of instructions between the branch and the merge
point, which may be useful for some control independence strategies.
The independent register set is the set of architectural registers that are
predicted to be independent of the branch. Post merge point instructions
that source registers identified by the independent register set do not have
any data-dependencies with instructions between the branch and its merge point.

\subsection{Merge Predictor Design} \label{sec:impl}

\begin{figure*}[t!]
  \centering
  \includegraphics[width=5.5in] {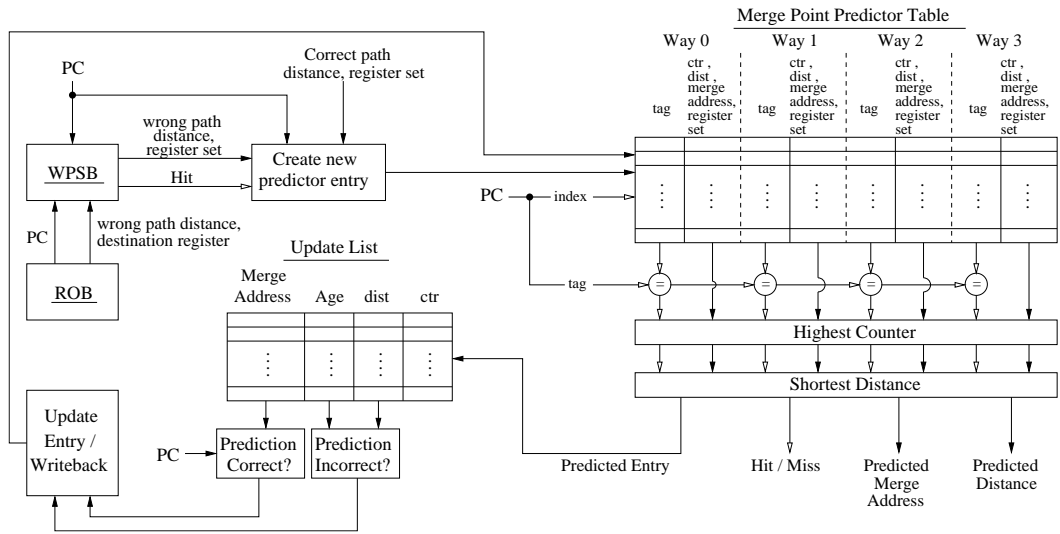}
  \caption{All three newly added structures: Merge Predictor Table, Update
  List and WPB.}
  \label{fig:monster}
  \medskip
  \raggedright
  \small
  \noindent
  \textbf{Figure \ref{fig:monster} shows the interactions between the three
    newly added structures and the ROB. The Predictor Table supplies the
    predicted entry to the Update List. The Update List compares entires to
    retired PCs until the merge point is confirmed or the merge distance is
    reached. At this point, the entry is updated and written back to the
    predictor. The WPB saves wrong-path PCs, supplied by the ROB, and
    compares them to correct-path PCs. When a match is found a new entry is
    installed in the predictor.}
\end{figure*}

The merge predictor design consists of three new structures: the Merge Point
Predictor Table, the Update List, and the Wrong Path Buffer (WPB). Figure
\ref{fig:monster} shows a block diagram of all three structures. The WPB is
responsible for detecting new merge points and installing them into the
predictor table. The update list is responsible for tracking predicted
entries and updating them appropriately.

Merge points are detected by observing both the wrong-path and correct-path
of a branch. When a branch misprediction occurs, wrong-path instructions are
copied from the Reorder Buffer (ROB) to the WPB. After the machine is
flushed, each retired, correct-path instruction accesses the WPB. If there
is a hit, then a new merge point has been found and is installed into the
predictor table.  Next time the branch is fetched, the predictor table
supplies the merge point and an entry is allocated in the update list. When
the branch retires, it activates its entry in the update list. Once
activated, the update list entry monitors retiring instructions. If the
predicted merge point retires within the merge distance without any
unexpected register writes, then the prediction is correct, otherwise it is
incorrect. In either case, the entry is updated and then removed from the
update list.

Section \ref{sec:impl:detect} discusses how new merge point are detected.
Next, \ref{sec:wpb} discusses the implementation of the WPB.  Section
\ref{sec:impl:pred} discusses how predictions are made.  Finally, section
\ref{sec:impl:update} discusses how the predictor is updated.

\subsubsection{Detecting New Merge Points} \label{sec:impl:detect} 
\begin{figure}[t!]
  \centering
  \includegraphics[width=3.5in] {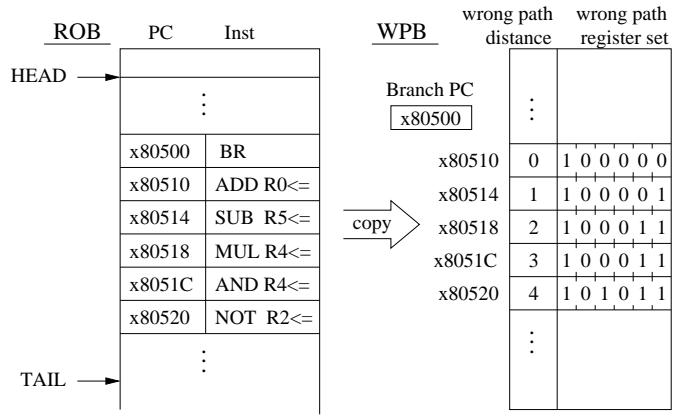}
  \caption{Wrong-path instructions copied from the ROB to the WPB}
  \label{fig:wpb}
  \medskip
  \raggedright
  \small
  \noindent
  \textbf{When the branch (PC=x80500) misprediction is detected, subsequent
  instructions are copied from the ROB to the WPB. The distance between the
  branch and each instruction is saved in the WPB. Additionally, the
  destination register of each instruction is saved in the the wrong-path
  register set bit-vector.}
\end{figure}

Our
design detects new merge points by exploiting branch mispredictions. Due to
the large size of instruction windows and high fetch rates, it is
common for processors to fetch many wrong-path instructions before
detecting a misprediction. Our experiments show an average of 100 dynamic
instructions fetched on the wrong path. Upon detecting a branch misprediction, wrong
path instructions must be copied from the ROB into the WPB. Figure
\ref{fig:wpb} shows an example. Instructions are copied from the ROB
starting with the first instruction after the mispredicted branch, and
ending upon one of three conditions: (1) there are no more instructions in
the ROB, (2) the maximum merge distance is reached, or (3) another instance
of the same branch is encountered in the ROB (i.e., a loop back to the
branch).  Instructions are copied from the ROB to the WPB by conducting a
ROB-walk during the flush.\footnote{We do not expect the ROB-walk latency to
be an issue. It is not unusual for ROB-walks to be used during a flush to
restore the state of the speculative register alias table.  Furthermore,
latency of ROB-walks are typically hidden by the front-end as it refills the
pipeline.} Each wrong-path instruction indexes the WPB with its PC and
stores a wrong-path distance number and a bit-vector called the wrong-path
independent register set. The distance number represents the number of
dynamic instructions between the current instruction and the branch, while
the independent register set represents the accumulated destination
registers of each instruction up to this point. Finally, we tag the WPB with
the PC of the mispredicting branch, and set a valid bit, indicating that the
WPB should be compared to future retired instructions.

After populating the WPB and flushing the machine, fetch is redirected down
the correct path. When correct-path instructions retire, their program
counters are used to index into the all of the valid WPBs. Similar to filling the WPB, 
we continue until one of three conditions is met: (1) a PC hits in the WPB (2)
the maximum merge distance has been reached or (3) the PC is equal to the PC
of the mispredicted branch\footnote{This happens when there is a loop back
to the branch before encountering the merge point.}. If there is a match
(i.e., option 1), then we have found a merge point and install a new entry
into the predictor table. If either option 2 or 3 occurs before finding a
match, then we assume that there is no merge point and invalidate the WPB.
Each WPB maintains a count of correct-path instructions that have indexed it
called the correct path distance. Additionally, the WPB also tracks the
correct-path independent register set by accumulating the destination
registers of retired instructions into a bit vector.

Upon a WPB hit, the wrong-path distance and wrong-path independent register
set are read from the WPB. The merge point is identified by the PC that hit
in the WPB. The predicted distance is set to the larger of the wrong path
distance and the correct path distance. Finally, the independent register
set is formed by ORing the wrong-path bit vector and the correct-path bit
vector.  The entry is then installed into the predictor table and the WPB is
invalidated.

\subsubsection{Design of the WPB} \label{sec:wpb}
Ideally the WPB would be a fully associative CAM, however, large CAMs
are impractical. For that reason, we chose to implement the WPB as a
128-entry 4-way set associative cache. Organizing the WPB
as a cache instead of a CAM creates the possibility for an entry to be
evicted, creating false negatives. In our evaluation, we observed less
than 1\% false negative rate, which led to an almost negligible loss in
coverage. The WPB uses the LRU replacement policy.

\begin{figure}[t!]
  \centering
  \includegraphics[width=2.75in]{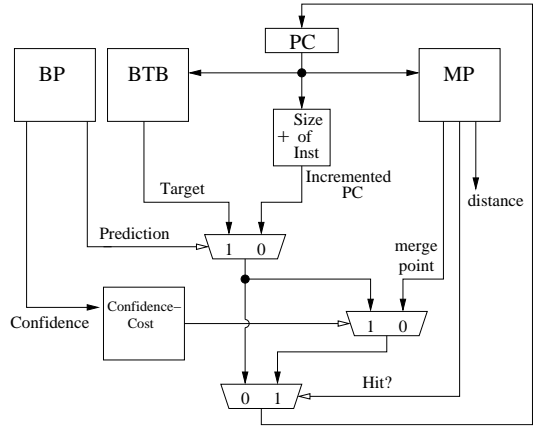} 
  \caption{Interaction between the Branch Predictor (BP), Branch Target
    Buffer (BTB), and Merge Predictor (MP).}
  \label{fig:front}
  \medskip
  \raggedright
  \small
  \noindent
\end{figure}

\subsubsection{Making the Prediction} \label{sec:impl:pred}
The PC is used to access the merge predictor in parallel with the branch
target buffer (BTB) and the branch predictor. Figure \ref{fig:front} shows
the connections between each of these structures. The entry supplied by the
merge predictor is only considered if the confidence-cost predictor has
identified the branch as hard-to-predict. The merge predictor is accessed as
a typical set associative cache. If there is a miss, we defer back to the branch
predictor. 

If there is a hit, it is possible that multiple entries match the
branch address. For example, consider the CFG in Figure \ref{fig:cfg}. As
discussed in section \ref{sec:motivation}, the merge point of the branch in
basic block A could either be D or F. It is possible that both D and F are
detected by the WPB and are both installed into the predictor. In the event
that two or more entries match in the predictor, the 3-bit saturating counter
is examined and the entry with the highest counter value is selected as the
prediction. If two or more entries have equal counter values, then the merge
entry containing the minimum merge distance is selected. We choose the entry
with the minimum merge distance because predicting smaller distances results in
smaller reservations in the instruction window. Once an entry has been
selected for prediction, all entires that matched in the predictor are
inserted into the Update List. 

When new entries are installed into the predictor, it may be necessary to
evict an older entry. Entries with the smallest counter value are the first
victims for eviction.  If all entries have equal counters, then the entry
with the largest predicted distance value is selected as the victim.

\subsubsection{Updating the Predictor} \label{sec:impl:update}

Once inserted into the update list, entries wait until the merge-predicted
branch instruction reaches retire. At that point, the update list entry
becomes active.  An Update List entry contains the following information:
(1) the PC of the merge-predicted branch, (2) a prediction age field, which
is the number of dynamic instructions retired since the entry became active,
and (3) the predictor table entry that will be updated and written
back to the predictor. An entry remains in the Update List until either (1)
a PC matching the merge address is found (meaning the prediction is
correct), (2) the age field exceeds the merge distance (the prediction is
incorrect), or (3) the PC of the merge-predicted branch is seen retiring for
a second time\footnote{This happens when there is a loop back to the branch
before encountering the merge point.} (the prediction is incorrect).  If the
prediction is correct, the 3-bit saturating counter is incremented,
otherwise the counter is decremented. Additionally, the destination
registers of gap instructions are monitored. If any unexpected writes
occur\footnote{Register writes not specified by the independent register
set.}, then the prediction is considered incorrect and the machine is
flushed. 

We introduce another update policy called UPDATE\_MAX. UPDATE\_MAX will not
remove an entry from the update list until the prediction age field has
exceeded the max prediction distance. In this mode, all entries in the
update list are treated as if their merge distances were equal to the max
merge distance, regardless of the actual prediction. This allows the update
list to detect if the merge address ever appears. If the merge address is
encountered, then the 3-bit prediction counter is incremented and the merge
distance field is set to equal the age field. This allows for the merge
distance field to be increased as necessary. If the max merge distance is
reached and the merge address is never found, then we decrement the 3-bit
counter, as before.

Once the update policy has completed, the entry is removed from the Update
List and written back to the predictor table. The Update List is a very small
table with only 8 entries, and thus is organized as a fully associative
cache. 
 % 2.5 pages
\begin{table*}[h]
  \centering
  \begin{tabular}{r|c|c|c|}
    \multicolumn{1}{c}{} & \multicolumn{1}{c}{} &
    \multicolumn{1}{c}{} & \multicolumn{1}{c}{} \\
    \multicolumn{1}{c}{} &
    \multicolumn{1}{c}{Low-Conf} & \multicolumn{1}{c}{Med-Conf} &
    \multicolumn{1}{c}{High-Conf} \\
    \cline{2-4}

    Lat-Low  & \textbf{MP} & BP & BP \\
    \cline{2-4}
    Lat-High & \textbf{MP} & \textbf{MP} & BP \\
    \cline{2-4}

  \end{tabular}
  \caption{Merge Point Prediction Decision Logic}
  \label{table:cc_decision_logic}
  \vspace{-0.1in}
\end{table*}

\section{Confidence-Cost Prediction} \label{sec:cc}

The confidence-cost predictor identifies branches that are likely to cause
expensive mispredictions. Our work focuses on improving all properties of
branch mispredictions that cause significant loss in fetch bandwidth. We
identify two such properties: (1) misprediction frequency and (2)
misprediction resolve latency. The first property is a measure of how often
branch mispredictions occur, while the second is a measure of wasted fetch
bandwidth per misprediction. Expensive branch mispredictions can result from
the extreme cases of either or both of these properties.  To our knowledge,
this work is the first to recognize branch resolve latency as an important
factor for identifying expensive branch mispredictions.

The confidence-cost predictor tracks both properties on a per-branch basis.
Each branch is assigned a confidence and cost value. Confidence is a
measure of how accurately a branch has been predicted in the past, while
cost is a measure of a branches resolve latency \footnote{Measured as the
number of cycles between prediction and the end of execution.}.
The confidence-cost predictor is essential for identifying branches that
pose large bottlenecks to fetch bandwidth, making them suitable
candidates for merge point prediction. Branches with either low confidence
or high resolve latency can significantly effect fetch bandwidth and should
be merge point predicted. Table \ref{table:cc_decision_logic} summarises the
conditions in which merge point prediction will be used over branch
prediction. We categorize all branches into three confidence levels
(Conf-Low, Conf-Med, Conf-High), and two resolution latency levels (Lat-Low,
Lat-High). Branches that correspond to entries labeled MP use the
merge point predictor, while the remaining branches will continue to use the
branch predictor.

\subsection{Measuring Branch Confidence}

We categorize all branches into three confidence levels: high, medium, and
low. To detect low-confidence, the 3-bit counter supplied by the highest
matching table in TAGE \cite{TAGE} is examined. If the counter is in either
the weakly-taken or weakly-not-taken state, then the prediction is labelled
as Conf-Low. Our experiments show that TAGE is only 70\% accurate when
labelled as Conf-Low \footnote{Measured across the SPEC CPU2006 \cite{Spec} benchmark
suite.}. To detect high-confidence, we use the JRS predictor \cite{jrs}. A
prediction is marked as Conf-High when it is not Conf-Low and the JRS
predictor reports high confidence. Branches not labelled Conf-Low or
Conf-High are labelled Conf-Med.

\subsection{Measuring Branch Cost}

Branch resolution times can vary dramatically due to long latency
instructions along the critical path of the branch (e.g., loads that miss in
the d-cache), resulting in some branches taking hundreds of cycles to
resolve. Long branch resolution times cause significant performance loss
even for branches that are predicted accurately 95\% of the time. We
introduce a new table for tracking average branch latency, called the branch
latency table. The branch latency table calculates the running average
\footnote{$Average Latency = 0.9*(New Latency) + 0.1*(Old Latency)$} of
each branch's resolve time. Branches with an average latency above the
threshold are labelled Lat-High, while the remaining branches are labelled
Lat-Low. For this work, we set the threshold value to be 50 cycles. % 1 page
\begin{table*}[t]
  \caption{System Configuration}
  \centering
  \begin{tabular}{|l|l|}
    \hline
    1: Core & 4-Wide Issue, 512-Entry ROB, 92-Entry Reservation Station,\\
            & TAGE Branch Predictor \cite{TAGE}, 3.2 GHz \\
    \hline

    \hline
    2: L1 Caches & 32 KB I-Cache, 32 KB D-Cache, 64 Byte Lines, 2 Ports, \\ 
                 & 3-Cycle Hit Latency, 8-Way, Write-Back.\\
    \hline
    3: L2 Cache & 1MB 8-Way, 18-Cycle Latency, Write-Back.\\
    \hline

    \hline
    4: Memory Controller & 64-Entry Memory Queue. \\
    \hline
    5: Prefetchers & Stream: 64 Streams, Distance 16. Prefetch into Last Level Cache.\\
    \hline
    6: DRAM & DDR3 \\
    \hline

    \hline
    7: Merge Predictor Table & 128 entiries, 4 way set associative, total size 1.6KB \\

    \hline
    8: WPB & 128 entries, 4 way set associative, total size 1KB \\

    \hline
    9: Update List & 8 entries, total size 113 bytes \\

    \hline
    10: Maximum Prediction Distance & 100 \\
    \hline
  \end{tabular}
  \label{table:scarab}
\end{table*}

\section{Evaluation Methodology} \label{sec:eval}
To simulate our proposal, we use a cycle-accurate x86 simulator. The
front-end of the simulator is based on Multi2Sim \cite{m2s}. The simulator
faithfully models core microarchitectural details and the cache hierarchy.
Table \ref{table:scarab} contains a
list of microarchitectural details. Our simulator includes a 64KB TAGE
\cite{TAGE} branch predictor configured similar to the version submitted to
CBP 2014. We did not include the SC or L components of the TAGE predictor.
We use the SPEC CPU2006 Integer benchmark suite \cite{Spec} to evaluate our
predictor. We use SimPoints \cite{Simpoints} methodology to identify
representative regions, and run all of our benchmarks for 200 million
instructions. 

We use several metrics to evaluate the effectiveness of our merge point
predictor. Accuracy is a measure of how often a prediction supplied by the
merge predictor is correct. Coverage is similar to accuracy, however it
factors in predictor misses (i.e., the merge predictor has no matching
entry).  Predicted distance is the predicted number of dynamic instructions
before encountering the merge point. True distance is the actual number of
dynamic instructions seen between the branch and merge point.  Finally, we
compute MPKI improvement, which is the difference between old MPKI and new
MPKI. New MPKI is calculated by adding the MPKI of the branch predictor and
the MPKI of the merge point predictor.

We evaluate two versions of the merge point predictor, a version that uses
the UPDATE\_MAX policy and a version that does not. We will refer to them as
MPPmax and MPP, respectively. Table \ref{table:scarab} shows the complete
predictor specification used in our experiments. We compare our predictor
against the infinitely sized reconvergence predictor introduced by Collins
et al.  \cite{Collins}. We will refer to their design as the
reconvergence-inf.

\begin{figure*}[t!]
  \centering
  \begin{subfigure}[b]{}
    \includegraphics[width=7.5in] {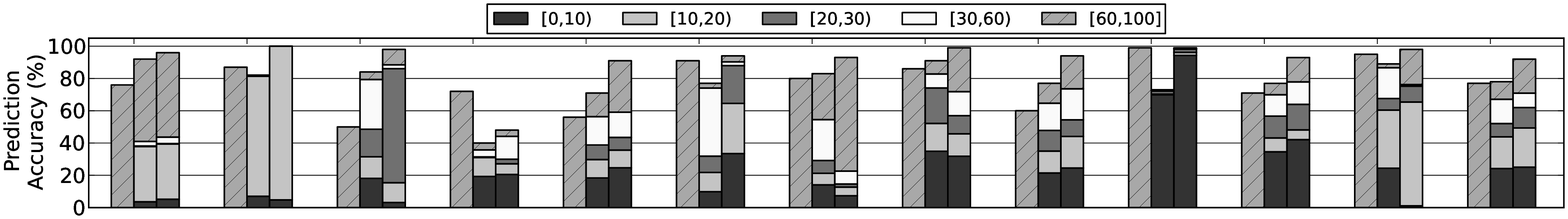}
    %\label{graph:accuracy_dist}
  \end{subfigure}

  \vspace{-0.06in}

  \begin{subfigure}[b]{}
    \includegraphics[width=7.5in] {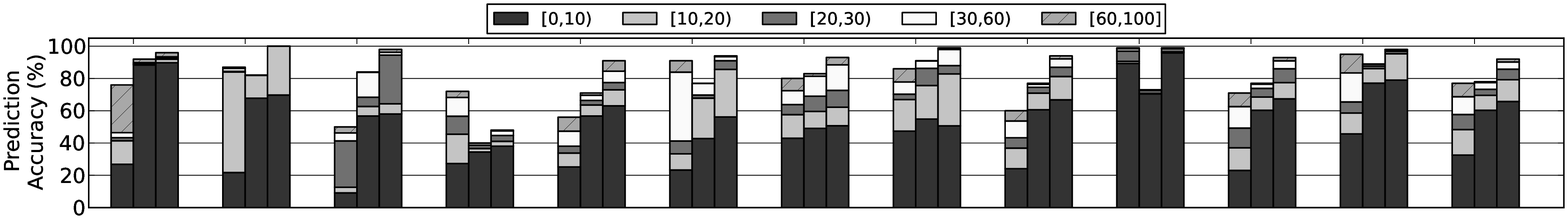}
    %\label{graph:accuracy_age}
  \end{subfigure}

  \vspace{-0.06in}

  \begin{subfigure}[b]{}
    \includegraphics[width=7.5in] {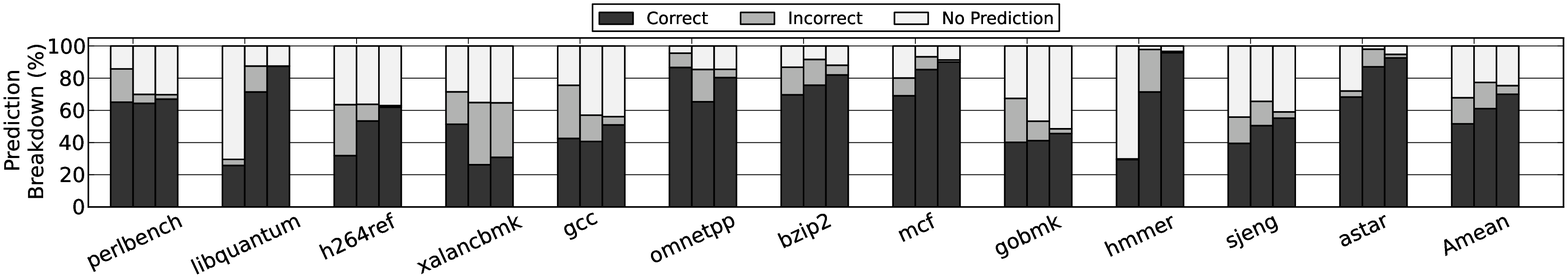}
    %\label{graph:coverage}
  \end{subfigure}

  \caption{ \textbf{
  The bars, from left to right, represent the infinitely sized
  reconvergence-inf\cite{Collins}, MPP, and MPPmax.
  The top graph (a) shows prediction accuracy overlayed with predicted distance. 
  The middle graph (b) shows prediction accuracy overlayed with true distance.
  The bottom graph (c) shows coverage.
  }}
  \label{graph:mp}
\end{figure*}

\section{Results and Analysis} \label{sec:results}

Figure \ref{graph:mp} (a) shows the accuracy of reconvergence-inf,
MPP, and MPPmax respectively. The height of the bars indicates accuracy,
while the stacks show predicted distances. Reconvergence-inf does
not predict distance, so we have shown all of its predicted distances as the
maximum distance. The final bar is the arithmetic mean (amean) of all
workloads. Figure \ref{graph:mp} (b) also shows prediction accuracy, but the
stacks show true distance values.  Figure \ref{graph:mp} (c) shows the
coverage results for each predictor.

\begin{figure}[t!]
  \centering
  \includegraphics[width=3.5in] {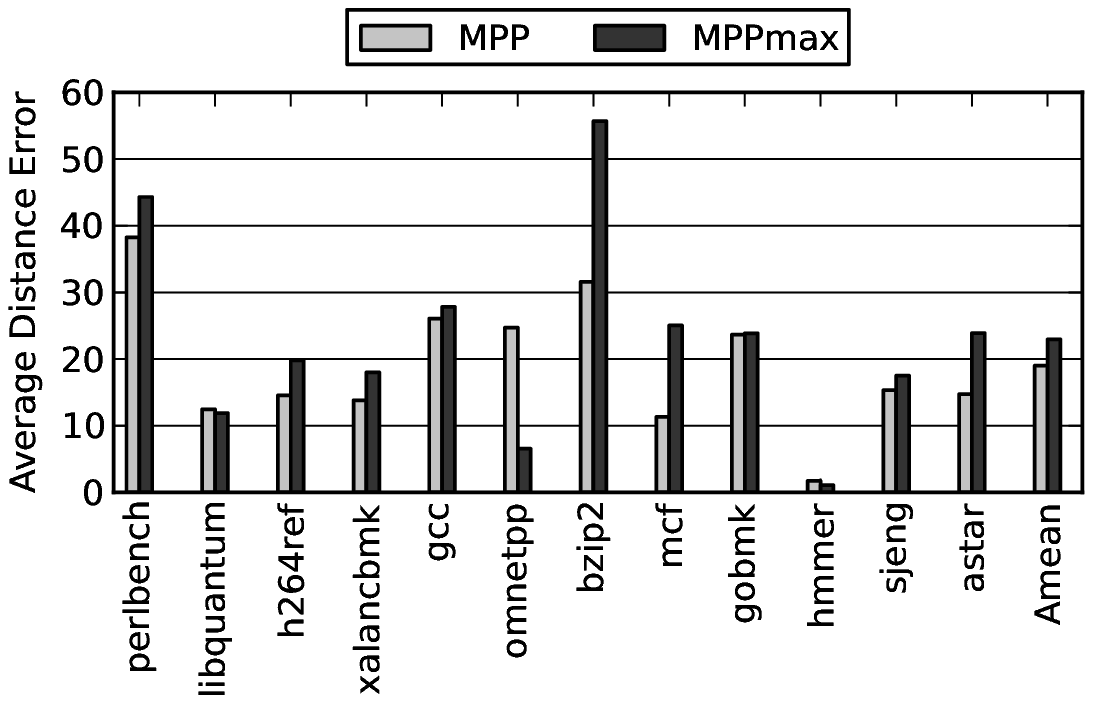}
  \caption{Average difference between predicted and true distances}
  \label{graph:agediff}
  \medskip
  \raggedright
  \small
  \noindent
\end{figure}

Ideally, the predicted distance and the true distance would be equal,
as some control independence strategies use distance to reserve space
in the instruction window. Unfortunately, this is not the case. Figure
\ref{graph:agediff} shows the average difference between predicted and true
distances for MPP and MPPmax. The height of each bar represents the wasted
space in the instruction window due to overestimating the predicted
distance.

Predicted distance can be overestimated for two reason. First, because the
predicted distance is the larger of the correct-path distance and the
wrong-path distance, it is possible that the smaller path was the one
actually traversed at runtime, thus creating an error. Shortening the error
in this case would be difficult, as the branch direction is not known. The
second case is that the update policy is installing an unnecessarily large
distance into the predictor. 

The accuracy of MPPmax is higher than MPP in every benchmark, resulting in
almost 14\% higher accuracy on average. This is because the UPDATE\_MAX
policy strictly increases the predicted distance over time. Predicting
larger distances can only increase prediction accuracy. However, the gain in
accuracy comes at a cost. The negative effects of UPDATE\_MAX are
shown in Figure \ref{graph:agediff}.  MPPmax overestimates distance to a
larger degree than MPP. This results in additional resources being wasted.

Both MPPmax and MPP outperform the infinitely-sized reconvergence-inf
predictor. Collins et al. \cite{Collins} work reports an accuracy of 95\% for
reconvergence-inf, however, our evaluation shows an accuracy of at most
78\%. We have accounted for the large discrepancy and attribute it to two
factors. First, we consider predictions incorrect once the predicted
distance has been exceeded. This is different from the methodology described
by Collins et al., however this difference does not lead to a significant
change in accuracy. Second, we enforce that all merge points identified by
both predictors are points where control actually converges. Due to the
methodology used by Collins et al., some of the predicted values are not
merge points.

\begin{figure}[t!]
  \centering
  \includegraphics[width=3.5in] {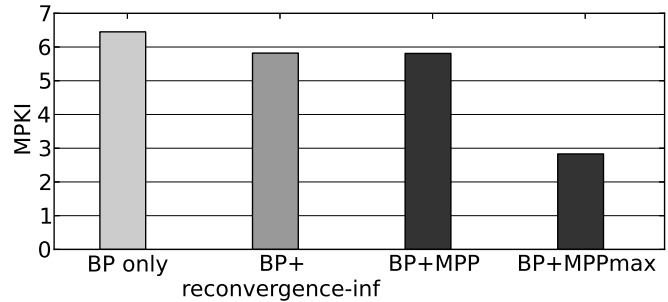}
  \caption{Projected MPKI with Merge Point Prediction}
  \label{graph:graph6}
  \medskip
  \raggedright
  \small
  \noindent
\end{figure}

Accuracy numbers alone are not enough to understand the worth of a merge
point predictor. It is important that we demonstrate that the merge point
predictor is accurate when the branch predictor is not.  Figure
\ref{graph:graph6} shows the difference in MPKI between a branch predictor
only design and a branch predictor + merge point predictor design. This
represents the total number of mispredictions created by both the branch
predictor and merge point predictor.  The figure shows that MPPmax is able
to achieve a 56\% improvement in MPKI over a BP only design and a 51\%
improvement over reconvergence-inf. This significant reduction in MPKI shows
that MPPmax is highly accurate, even in the presence of hard to predict
branches.

 % 0.5 pages
\section{Prior Work}

Control independence, as an alternative to branch prediction, was first
proposed by Lam and Wilson \cite{Lam}. While their study drew attention to
the area, it made many assumptions that led to an inaccurate upper bound on
performance \cite{Rotenberg, Sundararaman}. A more detailed analysis was
done by Rotenberg et al. \cite{Rotenberg}. They devised six models designed
to place an upper bound on the advantages of control independence. However,
the focus of their work was limited to theoretical benefits of control
independence, with no tangible solutions presented.

Skipper \cite{Cher} was the first microarchitecture proposal for control
independence that included a dynamic merge point predictor. Skipper
identified hard-to-predict branches, then fetched post merge point
instructions out-of-order to avoid prediction. However, Skipper is limited
to only predicting if-then, if-then-else, and loops. Furthermore, it makes
assumptions about the compiler and code layout, making it inaccurate and
resistant to change. Additionally, Skipper only uses branch confidence to
identify hard-to-predict branches. Our work also considers branch latency as
an important factor.

Collins et al. proposed a reconvergence predictor \cite{Collins}, which
attempted to solve the limitations of Skipper. Their algorithm introduced 3
different heuristics for detecting merge points of if-then and if-then-else
branches. Furthermore, they included support for call branches.  However,
their algorithm still depended on code layout, making it considerably
less accurate. Additionally, their prediction mechanism provided no support
for predicting distance or data-independence.  Our
predictor does not rely on the compiler, or make assumptions about code
layout. It can identify merge points of all branches with sufficiently low
distances and has a simply, extendable structure for tracking gap
properties. 

The SYRANT \cite{SYRANT} work symmetrically allocated resources along both
sides of a branch in preparation for a misprediction. Their merge point
prediction mechanism also used wrong-path information, however, the paper
focuses on the use case of their predictor, rather than evaluating the
prediction mechanism itself, leaving the design of their predictor largely
unknown. Additionally, their work is focused on optimizing branch
mispredictions, while our work is focuses on avoiding them all together. % 1 page
\section{Conclusion}
Improvements in branch prediction are not keeping up with the demand for
high fetch bandwidth. This results in branch mispredictions creating an
unacceptable gap in performance when compared to an oracle.  Exploiting
control independence is a promising alternative to branch prediction.  Our
work opens the door for control independence strategies to achieve high performance despite the existance of hard-to-predict branches. 
To accomplish this, we introduce a novel confidence-cost predictor that
identifies hard-to-predict and long latency branches. Our work is the first
to evaluate using branch resolve latency as an important factor for merge
point prediction. Second, we introduce a highly accurate dynamic merge point
predictor that produces better merge point predictions than a compiler and
prior work, achieving an accuracy of 95\%.
Together, these two techniques replace 56\% of all mispredicted branches with a correctly predicted merge point prediction. This result represents tremendous opportunity for control independence schemes, all while requiring no change to the software.
 % 1 page

\bibliography{ref}

\begin{thebibliography}{10}\setlength{\itemsep}{-1ex}\small

\bibitem{Spec}
The standard performance evaluation corporation (spec). the spec benchmark
  suite.

\bibitem{Al-Zawawi}
A.~S. Al-Zawawi, V.~K. Reddy, E.~Rotenberg, and H.~H. Akkary.
\newblock Transparent control independence (tci).
\newblock In {\em Proceedings of the 34th Annual International Symposium on
  Computer Architecture}, ISCA '07, pages 448--459, New York, NY, USA, 2007.
  ACM.

\bibitem{Cher}
C.-Y. Cher and T.~N. Vijaykumar.
\newblock Skipper: a microarchitecture for exploiting control-flow
  independence.
\newblock In {\em Proceedings. 34th ACM/IEEE International Symposium on
  Microarchitecture. MICRO-34}, pages 4--15, Dec 2001.

\bibitem{Chou}
Y.~Chou, J.~Fung, and J.~P. Shen.
\newblock Reducing branch misprediction penalties via dynamic control
  independence detection.
\newblock In {\em Proceedings of the 13th International Conference on
  Supercomputing}, ICS '99, pages 109--118, New York, NY, USA, 1999. ACM.

\bibitem{Collins}
J.~D. Collins, D.~M. Tullsen, and H.~Wang.
\newblock Control flow optimization via dynamic reconvergence prediction.
\newblock In {\em Microarchitecture, 2004. MICRO-37 2004. 37th International
  Symposium on}, pages 129--140, Dec 2004.

\bibitem{Hilton}
A.~D. Hilton and A.~Roth.
\newblock Ginger: Control independence using tag rewriting.
\newblock In {\em Proceedings of the 34th Annual International Symposium on
  Computer Architecture}, ISCA '07, pages 436--447, New York, NY, USA, 2007.
  ACM.

\bibitem{jrs}
E.~Jacobsen, E.~Rotenberg, and J.~E. Smith.
\newblock Assigning confidence to conditional branch predictions.
\newblock In {\em Proceedings of the 29th Annual ACM/IEEE International
  Symposium on Microarchitecture}, MICRO 29, pages 142--152, Washington, DC,
  USA, 1996. IEEE Computer Society.

\bibitem{Diverge-Merge}
H.~Kim, J.~A. Joao, O.~Mutlu, and Y.~N. Patt.
\newblock Diverge-merge processor (dmp): Dynamic predicated execution of
  complex control-flow graphs based on frequently executed paths.
\newblock In {\em Proceedings of the 39th Annual IEEE/ACM International
  Symposium on Microarchitecture}, MICRO 39, pages 53--64, Washington, DC, USA,
  2006. IEEE Computer Society.

\bibitem{Lam}
M.~S. Lam and R.~P. Wilson.
\newblock Limits of control flow on parallelism.
\newblock In {\em [1992] Proceedings the 19th Annual International Symposium on
  Computer Architecture}, pages 46--57, 1992.

\bibitem{Meng}
L.~Meng and S.~Oyanagi.
\newblock Control independence using dual renaming.
\newblock In {\em 2010 First International Conference on Networking and
  Computing}, pages 264--267, Nov 2010.

\bibitem{Simpoints}
E.~Perelman, G.~Hamerly, M.~Van~Biesbrouck, T.~Sherwood, and B.~Calder.
\newblock Using simpoint for accurate and efficient simulation.
\newblock In {\em Proceedings of the 2003 ACM SIGMETRICS International
  Conference on Measurement and Modeling of Computer Systems}, SIGMETRICS '03,
  pages 318--319, New York, NY, USA, 2003. ACM.

\bibitem{SYRANT}
N.~Premillieu and A.~Seznec.
\newblock Syrant: Symmetric resource allocation on not-taken and taken paths.
\newblock {\em ACM Trans. Archit. Code Optim.}, 8(4):43:1--43:20, Jan. 2012.

\bibitem{SPREPI}
N.~Pr{\'e}millieu and A.~Seznec.
\newblock {SPREPI: Selective Prediction and REplay for predicated
  Instructions}.
\newblock Research Report RR-8351, {INRIA}, Aug. 2013.

\bibitem{TraceProcessors}
E.~Rotenberg, Q.~Jacobson, Y.~Sazeides, and J.~Smith.
\newblock Trace processors.
\newblock In {\em Proceedings of the 30th Annual ACM/IEEE International
  Symposium on Microarchitecture}, MICRO 30, pages 138--148, Washington, DC,
  USA, 1997. IEEE Computer Society.

\bibitem{Rotenberg}
E.~Rotenberg, Q.~Jacobson, and J.~Smith.
\newblock A study of control independence in superscalar processors.
\newblock In {\em Proceedings Fifth International Symposium on High-Performance
  Computer Architecture}, pages 115--124, Jan 1999.

\bibitem{roth2002squash}
A.~Roth and G.~Sohi.
\newblock Squash reuse via a simplified implementation of register integration.
\newblock {\em Journal of Instruction-Level Parallelism}, 3:l--7, 2002.

\bibitem{TAGE}
A.~Seznec.
\newblock {TAGE-SC-L Branch Predictors}.
\newblock In {\em {JILP - Championship Branch Prediction}}, Minneapolis, United
  States, June 2014.

\bibitem{Sodani}
A.~Sodani and G.~S. Sohi.
\newblock Dynamic instruction reuse.
\newblock In {\em Proceedings of the 24th Annual International Symposium on
  Computer Architecture}, ISCA '97, pages 194--205, New York, NY, USA, 1997.
  ACM.

\bibitem{Sundararaman}
K.~K. Sundararaman and M.~Franklin.
\newblock Multiscalar execution along a single flow of control.
\newblock In {\em Proceedings of the 1997 International Conference on Parallel
  Processing (Cat. No.97TB100162)}, pages 106--113, Aug 1997.

\bibitem{m2s}
R.~Ubal, B.~Jang, P.~Mistry, D.~Schaa, and D.~Kaeli.
\newblock Multi2sim: A simulation framework for cpu-gpu computing.
\newblock In {\em Proceedings of the 21st International Conference on Parallel
  Architectures and Compilation Techniques}, PACT '12, pages 335--344, New
  York, NY, USA, 2012. ACM.

\end{thebibliography}
\bibliographystyle{latex8}

\end{document}